\documentstyle[12pt,amssymb,cite]{article}

\def\b{\begin{eqnarray}}
\def\e{\end{eqnarray}}
\def\l{\label}

\begin{document}

\title{INFLUENCE OF DIRECT URCA PROCESSES IN A STRONG
       MAGNETIC FIELD ON DYNAMICS OF COLLAPSING STAR ENVELOPE}

\author{
A.A.\,Gvozdev, I.S.\,Ognev\\[2mm]
\begin{tabular}{l}
Department of Theoretical Physics, Yaroslavl State University,\\
Sovietskaya 14, Yaroslavl 150000, Russia.\\
E-mail: {\em gvozdev@uniyar.ac.ru}, {\em ognev@uniyar.ac.ru}\\ [2mm]
\end{tabular}
}

\date{}

\maketitle

ABSTRACT. Direct URCA processes in a collapsing
star envelope with a strong magnetic field are investigated. It is shown
that in the toroidal magnetic field these processes can develope a torque
which quickly unwinds an envelope. A general expression for a force
density along the field direction is obtained. An influence of the neutrino
unwinding effect on a dynamics of the collapsar envelope is discussed
and numerically estimated. \\[1mm]


{\bf 1. Introduction}\\[1mm]


Collapsing star systems with millisecond remnants are of great interest in
astrophysics now. Such remnants can be produced, for example, in a type II
Supernova explosion \cite{BK1,W}, in a coalescense of a closed binary
system  of neutron stars \cite{J1} and  in an accretion induced collapse
\cite{Sp}. In the model of a spherically symmetric collapse, such a  systems
have common properties  during few seconds after a protostar contraction. At
this time a collapsing star remnant is formed. It is usually assumed that
the remnant consists of a compact rigid- rotating core and a differently
rotating envelope.

The compact core with the typical size $ R_o \sim 10 km $, the supranuclear
density $ \rho \gtrsim 10^{13} g/cm^3 $, and the high temperature
$ T \gtrsim 10 MeV $ is opaque to neutrinos. A remnant envelope with the
typical size of few tens of kilometers, the density
$ \rho \sim 10^{11} - 10^{12} g/cm^3 $ and  the temperature
$ T \sim 3 - 6 MeV $ is partially transparent to the neutrino flux.
Anomalously high neutrino flux with the typical luminosity
$ L_\nu \sim 10^{52} erg/s $ is emitted from the remnant during
2 - 3 seconds after the collapse. As the result of a high rotating frequency
and a medium viscosity of a remnant a turbulent dynamo and a large gradient
of angular velocities are inevitably produced during a star contraction.
The extremely strong poloidal magnetic field up to $ B \sim 10^{15} G $
could be generated by a dynamo process in a remnant\cite{Duncan}.
On the other hand, a large gradient of angular velocities in the vicinity
of a rigid rotating millisecond core can generate more strong toroidal
magnetic field (TMF) with the strength  $ B \sim 10^{15} - 10^{17} G $
during a second \cite{BK2}.
Note, that TMF can affect significantly the dynamics of the remnant
envelope even if this field exists during few seconds. For example, TMF
with $ B \sim 10^{17} G $ can initiate the process of a mantle shedding
\cite{BK1} and be an engine of an anisotropic $\gamma$-ray burst
\cite{W} in a Supernova II explosion.       \\[2mm]

{\bf 2. Momentum asymmetry in direct URCA processes}\\[1mm]

It is important to investigate the influence of the neutrino propagation
through a magnetized medium on dynamics of the remnant envelope. Indeed,
neutrinos are emitted and absorbed asymmetrically with respect to the
magnetic field direction \cite{Dorof} as the result of the parity
violation in weak processes. N.Chugai was the first who considered the
neutrino recoil momentum as a possible source of an anomalously large
pulsar kick velocity \cite{Ch}.

In the present paper we discuss an asymmetry of a momentum transferred by
neutrinos to the medium along the magnetic field direction in direct
URCA processes:
\b
p + e^- \Longleftrightarrow n + \nu_e ,
\l{1} \\
n + e^+ \Longleftrightarrow p + \tilde\nu_e
\l{2}
\e
The asymmetry can lead to a macroscopic torque spins up the envelope with
TMF.

A  quantitative estimation of the effect can be obtained from the
four-vector of the energy-momentum transferred by neutrinos to a unit volume
of the envelope per unit time:
\b
\frac{dP_\alpha}{dt} = \left( \frac{dQ}{dt}, \vec\Im \right) =
\frac{1}{V} \int \prod\limits_i dn_i f_i
\prod\limits_f dn_f (1 - f_f) \frac{|S_{if}|^2}{{\cal T}} k_\alpha ,
\l{dpa}
\e
where $dn_i$ and $dn_f$ are the initial and final state numbers in the phase
space element,$f_i$ and $f_f$ are the distribution functions of initial and
final particles, $k_\alpha$ is the neutrino four- momentum transferred in
a single reaction, $|S_{if}|^2 / {\cal T}$ is a process  S-matrix element
squared per unit time.

To calculate the components of the energy-momentum vector we note that,
in accordance with envelope conditions, the nucleonic gas is Boltzmann and
nonrelativistic one. We also assume that medium parameters and the
magnetic field strength satisfy following unequalities:
\b
m_p T \gg eB \gtrsim \mu_e^2, T^2 \gg m_e^2   .
\e
It means that ultrarelativistic electrons and positrons occupy the ground
Landay level only while the protons occupy quite many levels.
We employ the neutrino distribution function in absence of the magnetic
field. This is a good approximation when the region occupies by the
strong magnetic field is smaller or of the order of the neutrino mean-free
path. In the TMF generation model the region occupies by such a field is not
larger then several kilometers \cite{BK2}, while the estimation of the
neutrino mean- free path due to direct URCA processes gives:
\b
l_{\nu} \simeq 4 km
\left( \frac{ 4.4 \times 10^{16} G}{ B } \right)
\left( \frac{ 5 \times 10^{11} g / cm^3 }{ \rho } \right) \; . \nonumber
\e
So, the magnetic field cannot strongly influence the neutrino distribution
and we can use the one-dimensional factorized neutrino distribution
function:
\b
f_{\nu} = \frac{\Phi_{\nu}(r,\chi)}
{ \exp{(\omega / T_\nu - \eta_\nu)} + 1 }\ ,
\e
where $T_\nu$ is the neutrino spectral temperature, $\eta_\nu$ is a fitting
parameter, $\chi$ is the cosine of the angle between the radial direction
and the neutrino momentum.

Under all these assumptions we obtain the following general expression
for the force density along the magnetic field direction in URCA
processes:
\b
\Im _\| ^{(urca)} &=&
{\cal N}  \bigg( 3 \langle \chi^2 \rangle - 1 \bigg)
\frac{N_n}{N_B}
\Bigg[ \exp{( \delta \eta - a)} I(a) + I(-a) \Bigg] -
\nonumber \\
 &-& \; \frac{1}{2}  \frac{g_a^2 - 1}{3 g_a^2 + 1}
\bigg( 1 - \langle \chi^2 \rangle \bigg)
 \frac{dQ^{(urca)}}{dt}
\; , \l{F} \\
&&I(a) = \int\limits_0^\infty \frac{y^3 dy}{e^{y-a} + 1}, \;\;\;
a = \frac{\mu_e - (m_n - m_p)}{T} , \;\; \; \nonumber \\
&&\delta\eta = (\mu_e+\mu_p-\mu_n)/T , \;\;\; N_B = N_n + N_p,
\nonumber \\
&&\langle \chi^2 \rangle = \langle \chi^2_{\tilde\nu} \rangle
\simeq \langle \chi^2_{\nu} \rangle =
\left(
\int \chi^2 \;\omega \;f_{\nu} \;d^3k
\right)
\left(
\int \omega \;f_{\nu} \;d^3k
\right)^{-1} ,
\nonumber
\e
where $\mu_e$, $\mu_p$ and $\mu_n$ are chemical potentials of electrons,
protons and neutrons, $N_n$ and $N_p$ are the number densities of neutrons and
protons. The dimensional factor ${\cal N}$ is defined as:
\b
{\cal N} = \frac{G_F^2 \cos^2 \theta_c}{ (2\pi)^3 } \;
\frac{g_a^2 - 1}{3} \; eB \; T^4 \; N_B ,
\e
where $ g_a \simeq 1.26 $ is the axial constant of the nucleonic current, $G_F$
is the Fermi constant, $\theta_c$ is the Cabbibo angle. As it is seen from
Eq. (\ref{F}), the neutrino momentum asymmetry transferred to
the shell along the magnetic field is produced by two sources: the first one
is an anisotropy of the neutrino distribution
($ \langle \chi^2 \rangle \ne 1/3 $) and the second one is the energy transfer
to the shell by URCA processes ($ dQ^{(urca)}/dt \neq 0 $). Hence, the momentum
asymmetry takes place in the case of non-equilibrium neutrino distribution
function only \cite{Kusenko}. As for the envelope medium, it is considered to
be in the local quasi-equilibrium which is defined by the following conditions:
\b
\Gamma_{n \to p} = \Gamma_{p \to n}, \;\;\; dQ^{(urca)} / dt = 0, \nonumber
\e
where $\Gamma$ is the reaction rate. We assume the medium density and the
field strength to be preset values. Using the additional conditions:
$ d\rho / dt = 0$ and $ N_p = N_{e^-} - N_{e^+}$, we can obtain all medium
parameters from the conditions of the local quasi-equilibrium.
\\[2mm]

{\bf 3. Numerical estimations}\\ [1mm]
For numerical estimations we use neutrino parameters:
$ T_{\nu_e} \simeq 4 MeV $, $ T_{\tilde\nu_e} \simeq 5 MeV $ and
$ \eta_\nu \simeq \eta_{\tilde\nu} \simeq 0 $ from Ref. \citen{J2}, where
the Boltzmann equation in the model of a spherically symmetric collapse
on the basic neutrino emission stage was solved numerically. For the
envelope region with the typical density $ \rho = 5 \times 10^{11} g/cm^3 $
the numerical value of the mean cosine squared is
$ \langle \chi^2_{\nu} \rangle \simeq 0.4 $. Provided that the magnetic
field strength  is $ B = 4 \times 10^{16} G $, we obtain from the conditions
of the local  quasi-equilibrium:
\b
T \simeq T_{\nu_e}, \;\;\; a \simeq 3, \;\;\; N_p / N_B \simeq  0.07 .
\nonumber
\e
Using these parameters, we estimate the force density along the magnetic field
direction (\ref{F}), as:
\b
\Im^{(urca)}_{\|} \simeq 1.8 \times 10^{20}  dynes / cm^3
\left( \frac{T}{4 \; MeV} \right)^4
\left( \frac{B}{ 4.4 \times 10^{16} G } \right)
\left( \frac{ \rho }{ 5 \times 10^{11} g /cm^3 } \right) \; .
\nonumber
\e
As one can see, the force density  directs along the magnetic field
$ \vec B $.
Surprisingly, another important process of neutrino-nucleon scattering
gives the contribution to the force density of the same sign and
approximately twice larger than the direct URCA \cite{OG}.
Hence, the estimation for the total force density in the dominant processes
(the direct URCA and the neutrino-nucleon scattering) is:
\b
&&\Im^{(urca)}_{\|} \simeq 5 \times 10^{20}  dynes / cm^3
\left( \frac{B}{ 4.4 \times 10^{16} G } \right)
\left( \frac{ \rho }{ 5 \times 10^{11} g /cm^3 } \right) \; .
\nonumber
\e
The angular acceleration produced by the torque exerts by such a force is:
\begin{eqnarray}
\dot \Omega \sim 10^{3} s^{-2}
\left( \frac{B}{ 4.4 \times 10^{16} G } \right)
\left( \frac{ R_o }{ 10 km } \right).
\nonumber
\end{eqnarray}
Note that this acceleration is large enough to spin up the region of
the remnant envelope containing the strong TMF to the typical velocities of
millisecond pulsars in a time of order a second.
\\[2mm]

{\bf 4. Conclusion}\\ [1mm]

We consider collapsing star systems with millisecond remnants containing
a strong TMF on the stage of a basic neutrino emission and estimate the force
density along the magnetic field direction in direct URCA  processes.
The angular acceleration produced by the force is large enough to affect
significantly on the remnant magneto-hydro-dynamics (MHD).
Hence, the neutrino "spin-up" effect should be taken into account in analysis
of the full system of MHD equations. In particular,  this effect can
influence on the mechanism of the TMF generation, the mantle-shedding
process, the mechanism leading to the formation of anisotropic GRB and
also can provoke a formation of some MHD instabilities. \\[3mm]

{\it Acknowledgements.} The authors express the deep gratitude to the
Organizing Committee of the GMIC' 99 Conference for the possibility to
participate and for the worm hospitality. This work was
supported in part by the INTAS under Grant No. 96-0659 and by the Russian
Foundation for Basic Research under Grant No. 98-02-16694.
\\[3mm]
\indent
{\bf References\\[2mm]}

\vfill

\end{document}